# Noctilucent Clouds Polarimetry:
# Twilight Measurements in a Wide Range of Scattering Angles

Oleg S. Ugolnikov[1], Igor A. Maslov[1], Boris V. Kozelov[2], Zhanna M. Dlugach[3]

[1]Space Research Institute, Russian Academy of Sciences,
Profsouyznaya st., 84/32, Moscow 117997 Russia
[2]Polar Geophysical Institute, Akademgorodok str., 26a, Apatity 184209 Russia
[3]Main Astronomical Observatory, National Ukrainian Academy of Sciences,
Akad. Zabolotnogo st., 27, Kiev 03680 Ukraine

Corresponding author e-mail: ougolnikov@gmail.com

**Abstract**
Wide-field polarization measurements of the twilight sky background during the several nights with bright and extended noctilucent clouds in central and northern Russia in 2014 and 2015 are used to build the phase dependence of degree of polarization of sunlight scattered by clouds particles in a wide range of scattering angles (from 40° to 130°). This range covers the linear polarization maximum near 90° and large-angle slope of the curve which are most sensitive to the particle size. The method of separation of scattering on clouds particles on the twilight background is presented. Results are compared with T-matrix simulations for different sizes and shapes of ice particles, the best-fit model radius of particles (0.06 μm) and maximum radius (about 0.1 μm) are estimated.

## 1. Introduction

Noctilucent clouds (NLCs) are well-known to be the highest clouds in the Earth's atmosphere. They appear in the summer mesosphere as a thin layer. Altitude measurements conducted over more than 100 years and listed in (Bronsten and Grishin, 1970; Gadsden and Schröder, 1989) give good agreement of between 82 and 84 km. These clouds are also the youngest (or the latest observed) in the whole atmosphere, having first been described by Leslie (1885).

NLCs are known to consist mainly of water ice (Hervig et al., 2001). For ice crystals to be created and to grow so high above the ground, very low temperatures together with a significant amount of water vapor are required. These factors could have changed during the last century, leading to an increase in the size of crystals and to the appearance of visible NLCs. Human activity can influence this process not only by contributing an increase in the quantity of $H_2O$, but also of $CH_4$ producing $H_2O$ molecules in the mesosphere (Thomas and Olivero, 2001) and of $CO_2$ causing the radiative cooling effect (Roble and Dickinson, 1989). For this reason NLCs are a useful tool for studying global climate changes in the atmosphere (Berger and Lübken, 2015; DeLand and Thomas, 2015). However, no changes in NLCs characteristics were observed during the last decades (Romejko et al. 2003; Dalin et al., 2006; Pertsev et al., 2014).

Wave effects and global atmospheric circulation lead to an inverted annual temperature cycle in mid-latitude mesosphere with summer minimum. Temperature values above 80 km can be lower than 150 K. For present $H_2O$ concentration in the mesosphere, this temperature corresponds to the saturation pressure of water vapor and seems to be a threshold for NLCs to be observed (Dalin et al., 2011). These conditions may occur from late May until early August between latitudes +50° and +60°. Fortunately, the upper mesosphere can remain sunlit during most of or even the whole of the night, providing excellent conditions for observation of NLCs. In latitudes northwards of +60° summer nights in June and July are too bright to be able to see NLCs, but lower temperatures have resulted in the duration of NLC activity to expand to May and August, when they can be observed.

Satellite observations show the presence of semi-permanent clouds around the poles (Donahue et al., 1972; Donahue and Guenther, 1973; Russell et al., 2009).

However, the mechanism of NLC particle formation has no complete explanation yet. Most probably, their condensation nuclei are particles of extraterrestrial origin (Rosinski and Snow, 1961). Meteoroids entering the Earth's atmosphere are of size larger than several microns; smaller particles are blown out of the Solar System by radiation pressure (Soberman, 1969). In major meteor showers, the particles have orbits close to that of their parent comet, and therefore, the value of solar radiation pressure should be less compared with the gravitational attraction experienced by these bodies. Tiny particles of about 1 μm for Earth-encountering showers and dozens of microns with sporadic velocities can enter the atmosphere without ablation or fragmentation (Whipple, 1950), revealing themselves by weakly-polarized scattering of light during twilight (Ugolnikov and Maslov, 2007, 2014).

Larger meteoroids crumble into smaller particles of submicron sizes have been collected after meteor showers maxima such as the outburst of Leonids in 1965 (Skrivanek et al., 1969; Hemenway and Hallgren, 1969). Bodies with even larger sizes vaporize, yielding metal-rich material which then recondenses to nanometer-sized particles known as "meteor smoke" (Hunten et al., 1980), those can provide condensation nuclei for NLCs.

Another possible condensation mechanism is related to the hydrated ions in the mesosphere (Witt, 1969). Theoretical modeling of this mechanism for NLC particle formation was performed by Turco et al. (1982) and subsequently improved in the CARMA model (e.g. Rapp and Thomas, 2006). The model calculations show that the ice particles grow in a layer close to the summer mesopause (from 80 to 90 km), but the largest particles of size about 0.1 μm are drifted to the lower boundary of this layer. Moving further downwards, the ice particles sublimate. This explains why the NLCs layer is so thin.

The basic tool for studying the microphysical characteristics of NLCs particles as the large-size limit of polar mesospheric clouds particles (Gadsden and Schröder, 1989) is optical remote sounding. The review of Kokhanovsky (2005) describes various results obtained by different methods. Rocket-borne UV-photometry (Gumbel et al., 2001; Gumbel and Witt, 2001) had shown that the typical sizes of NLCs particles are less than 50 nm, based on an assumption that their shape is spherical. Estimates using multi-color lidar sounding (Von Cossart et al., 1999; Alpers et al., 2000; Baumgarten et al., 2007) and spaceborne spectroscopy (Carbary et al., 2002; Von Savigny et al., 2004; Von Savigny and Burrows, 2007) produce similar values for the sizes of NLC particles.

Since the first rocket sounding experiments, it became obvious that the most sensitive test for the shape of the particles is polarization study (Gadsden and Schröder, 1989). The best confirmation is the detection of depolarization effect by cross-polarization lidar sounding (Baumgarten et al., 2002) that can be observed only in the case of non-spherical particles. The same can be confirmed by the discovery of circular polarization of light scattered by NLCs (Gadsden et al., 1979). However, as noted there, it could be also the effect of scattering of initially polarized atmospheric background, or, in other words, the multiple scattering in the atmosphere with participation of NLCs particles. We shall discuss this effect in more details in this paper.

The first measurements of linear polarization of the light scattered by NLCs were carried out from the ground (Witt, 1957) and rockets (Witt, 1960). The rocket measurements were conducted in the visible spectral range (490 and 610 nm) for scattering angles θ up to 80°. The observed degree of polarization was slightly less than the Rayleigh level, that corresponded to a value of about 130 nm for the particle radius. The polarization measurements were continued by Vasilyev (1959, 1962), Tarasova (1962), and Willmann (1962). The degree of polarization for small scattering angles was low or even negative (Vasilyev, 1959). This led to significant overestimation of particle radius (up to 500-700 nm), the reason of such an effect will be discussed below. Willmann (1962) had found more reliable value for the particle radius, 140 nm. The same result for wavelength 540 nm and θ around 80° was obtained later by Tozer and Beeson (1974). The measurements at shorter wavelengths in optical and UV-ranges (Witt et al., 1976; Heitzenberg et al., 1978) had also shown strong linear polarization at the same angle that reduced the size of NLCs particles.

All the above-mentioned polarization measurements were made in a range of small scattering angles. This is natural since NLCs are observed mostly in the dawn/dusk sky area. As was shown by (Gumbel et al., 2001), the particles with radii as small as quarter of wavelength do already have a significant excess of scattered emission at small angles θ, and scattered light is fainter at θ>90°. Observing from the ground at twilight yields another problem: the altitude of the Earth's shadow near the zenith and in the dawn-opposite sky area rises fast, and NLCs can be visible only during the lighter twilight stage, at lower solar depressions below the horizon. This means higher level of sky background consisting basically of multiple scattering in troposphere (Ugolnikov, 1999; Ugolnikov and Maslov, 2002).

However, it is polarization measurements in the range θ>90° which are the most interesting and sensitive to the particles size and shape. Figure 1 shows the theoretical dependencies of polarization (calculated as the ratio of two parameters of the Mie scattering matrix $-F_{12}/F_{11}$) for ice spheres (following Iwabuchi and Yang, 2011, we assume refractive index 1.31, the question is discussed below) with radii from 0.1 to 0.2 μm. For small scattering angles the curves are close to each other, while the polarization slowly decreases with the particle size. The same effect can be caused by non-subtracted background or multiple scattering that will not disappear even if the observer is located outside the atmosphere. NLCs will still scatter the emission of atmospheric background. This can lead to overestimation of NLC particle size.

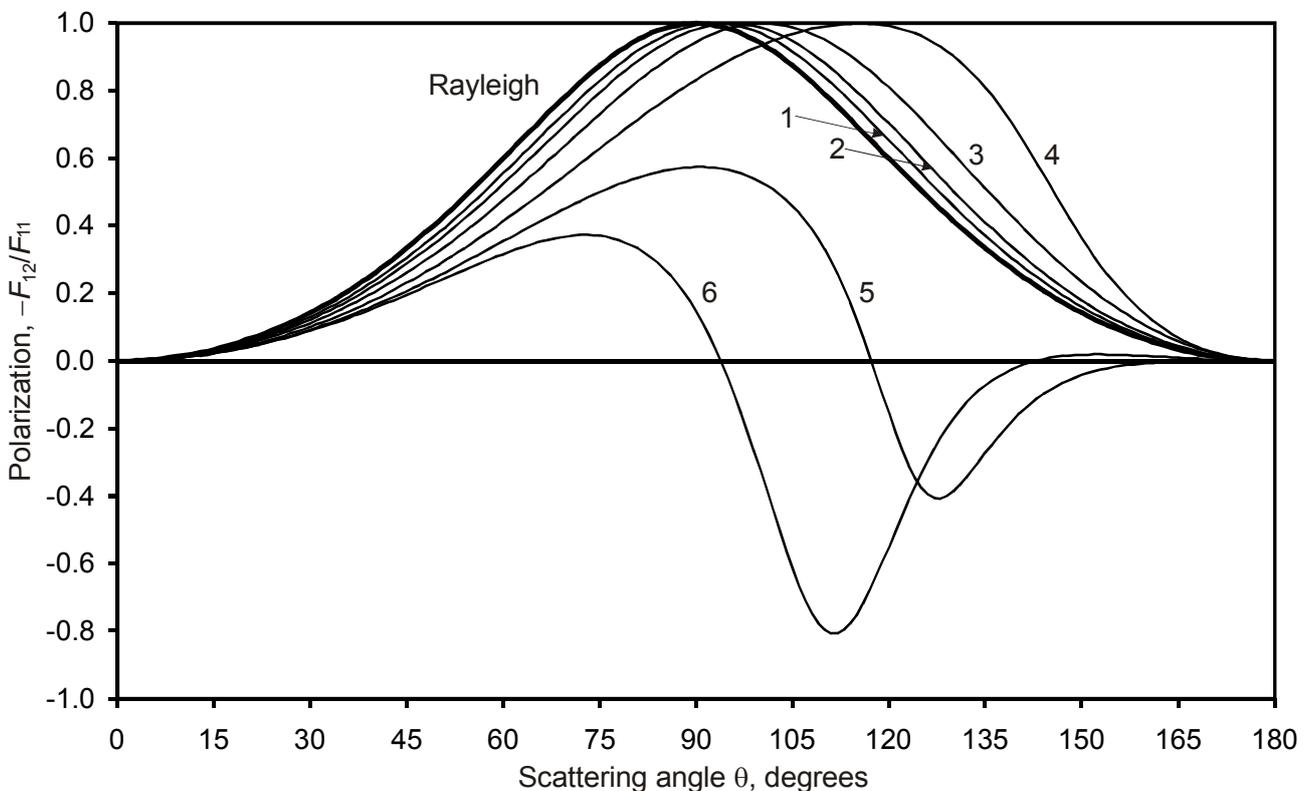

*Figure 1. Theoretical dependencies of polarization on the scattering angle for ice spheres with different radii (bold line – the Rayleigh limit, 1 – 0.10 μm; 2 – 0.12 μm; 3 – 0.14 μm; 4 – 0.16 μm; 5 – 0.18 μm; 6 – 0.20 μm).*

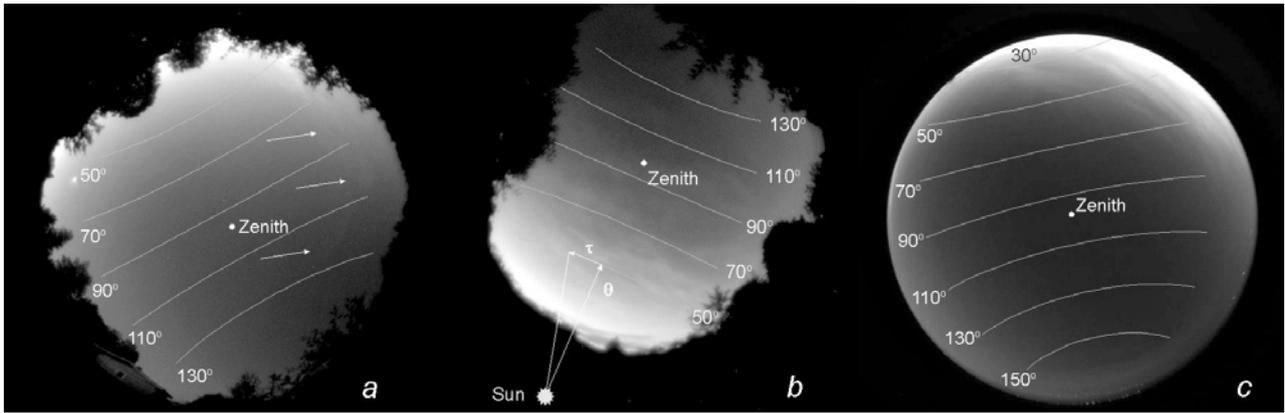

*Figure 2. Sky images with noctilucent clouds in the morning of July 22, 2014, clouds are shown by arrows (a), in the evening of July 5, 2015 (b), and in the evening of August 15, 2015 (c). Solar zenith angle is close to 97°. Images are processed to increase the contrast of the clouds. The coordinate system and arcs of constant scattering angles are shown.*

The situation is principally different for higher scattering angles: initially the degree of polarization rises, while the position of maximum is shifting from 90° to the larger angles (Rayleigh-Gans approximation case), but then the degree of polarization rapidly falls. Presence of large particles (more than 0.1 μm by radius) brings the asymmetry to the measured polarization distribution around θ=90° that can be measured. However, non-spherical particles have different polarization dependencies that will be taken into account in this paper.

## 2. Observations

In our work we use the twilight background survey by all-sky cameras designed for polarization measurements. The first device is Wide-Angle Polarization Camera (WAPC) had started to operate in Chepelevo, central Russia (55.2°N, 37.5°E) in summer 2011. The camera is described in (Ugolnikov and Maslov, 2013ab) and allows to build the maps of intensity and polarization (three Stokes vector components) of the twilight sky with zenith distances up to 65° (however, this value is reduced in some directions due to landscape restrictions). The diameter of the sky image is about 500 pixels. The exposure time depends on the twilight stage and vary from 1 to 15 seconds during the transitive and dark twilight, when NLCs are observed.

The main goal of this project is thermal analysis of the mesosphere (Ugolnikov and Maslov, 2013ab) and detection of dust (Ugolnikov and Maslov, 2014). The spectral band is wide (FHWM around 80 nm), the effective wavelength is 540 nm, almost the same as in rocket-borne NLCs polarimetry (Tozer and Beeson, 1974) and cross-polarization lidar sounding (Baumgarten et al., 2002). The observations are hold mainly in summer. NLCs were visible several times, but usually they were seen not higher than 20-25° above the northern horizon at small scattering angles.

The second camera is installed in Apatity, northern Russia (67.6°N, 33.4°E) and has been operating by the same program since early 2015 (Ugolnikov and Kozelov, 2015). The spectral band is close to that of the first camera (effective wavelength 530 nm, FHWM about 70 nm), with field of view 180°. We do not use the data with zenith angles more than 65° in order to avoid the effects of light pollution and tropospheric aerosol. The camera stops operating during the season of midnight sun and light twilight from early May till mid-August, but can register NLCs just before and after this interval. The technical characteristics of the observations are close to the ones of the first camera.

To study the polarization properties of light scattered by NLCs from the ground, it is highly preferable that the clouds cover the area with a wide range of scattering angles or (better) the major part of the sky including the zenith. The first wide-ranging appearance of NLCs in Chepelevo occurred during the morning of July 22, 2014, see Figure 2a. The clouds were seen far to the left of the Sun, achieving scattering angles of about 120°. However, they were not so bright, especially at scattering angles of around 90° (NLCs are shown by arrows in Figure 2a).

The rare case of all-sky covering NLCs occurred at the same location during observations in the evening of July 5, 2015. Clouds were bright and visible within a wide range of scattering angles up to 130° (Figure 2b), providing the best conditions for measuring the scattered radiation field. The major part of the sky higher than 30° above the horizon was totally free from tropospheric clouds during a short interval with solar zenith angles of about 97°-98°. This is the twilight stage characterized by maximum contrast of NLCs in the sky. The data of this evening is the primary for our study of the microphysical properties of NLCs particles. The data for other twilight observations are used for comparison with the basic one to check the accuracy of separation of the light scattered by NLCs in the case of fainter clouds and estimate the influence of twilight background and multiple scattering.

Finally, bright NLCs at scattering angles of up to 70° were observed in Apatity just during the first night measurements after the summer break, August 15, 2015 (Figure 2c). All these twilights were moonless except for the morning of July 22, 2014, when the crescent Moon was situated far from NLCs (it is seen in Figure 2a). The contribution of moonlight scattering to the twilight background was small. To fix this, we do not take into account the Moon-surrounding part of the sky during the data processing.

Since we are mainly interested in the angle dependence of the degree of polarization of light scattered by NLCs, we use a different coordinate system during data processing than the one in (Ugolnikov, Maslov, 2013b). This is the polar coordinate system shown in one WAPC image with NLCs in Figure 2b. One of the poles coincides with the position of the Sun corrected by a small refraction angle for tangent upper tropospheric rays (about 0.2°). The brightness and polarization were measured at sky points along the arcs of minor circles with fixed angular distances from the Sun or, equivalently, constant single scattering angles θ (Figure 2b) from 30° to 130°. The azimuth angle τ (Figure 2b) varied from –60° to +60°, if there were no additional landscape restrictions. The data were binned inside circles with radii 1° and processed with angular resolution 2°. Meridian major semi-circle connecting the Sun and the observation point is the basic coordinate line for Stokes vector calculations.

## 3. Temperature conditions

In the cases of NLCs observations we can start from the temperature analysis. Gadsden and Schröder (1989) showed that NLCs appear mainly in the cold periods, however, there is no significant numerical correlation between NLC frequency and the temperature. Dalin et al. (2011) point to the temperature slightly below 150 K as necessary but not enough condition of NLCs appearance.

We study the relation of bright NLCs appearance with temperature at the same altitude (83 km) not only at the same moment; we also check its evolution during the previous days. For this purpose, we have taken the EOS Aura/MLS satellite data (Schwatrz et al., 2008; EOS MLS Science Team, 2011) averaged over one satellite revolution by the nearby locations (±3° by latitude, ±10° by longitude from the observation place). These values are shown in Figure 3. For July 2014, when WAPC dark twilight and night background data is also available, we use the method (Ugolnikov and Maslov, 2013b) and also plot the temperature values at 83 km obtained by the same technique.

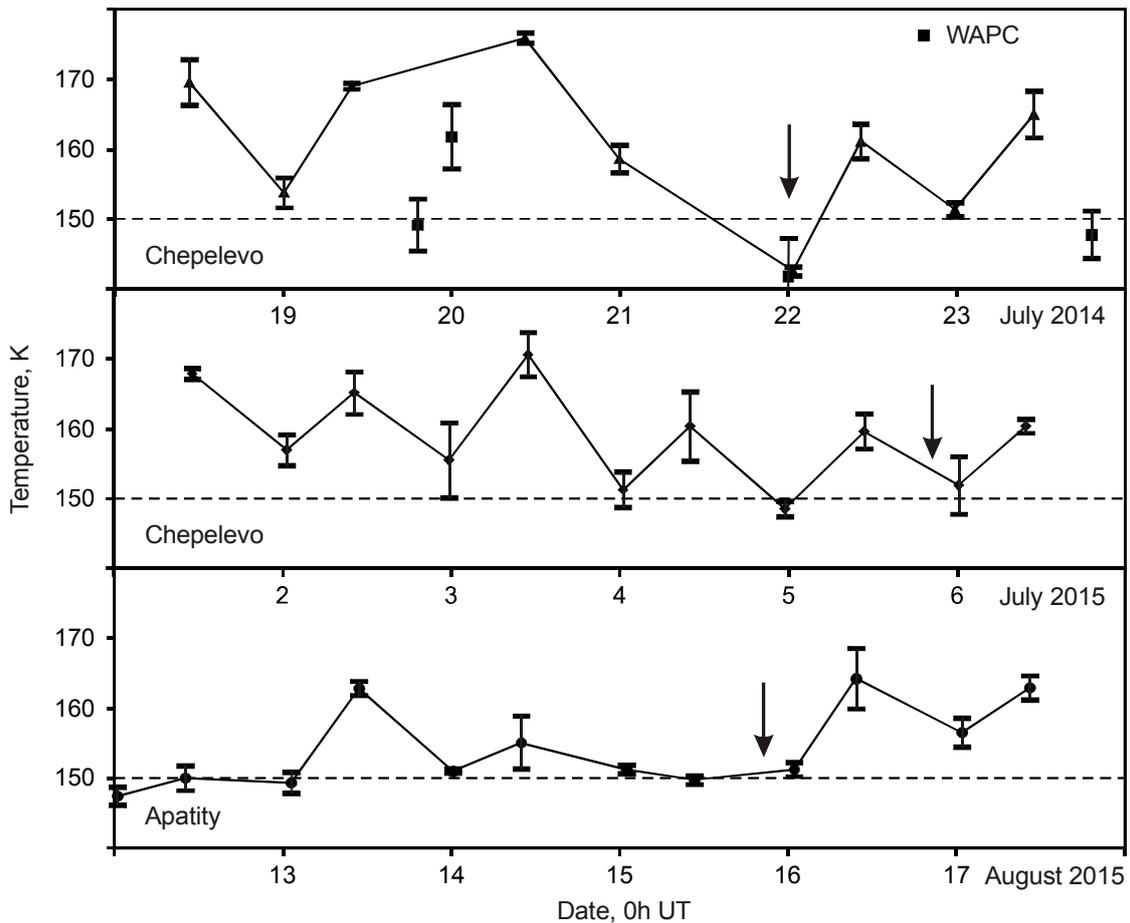

*Figure 3. EOS Aura/MLS and WAPC temperatures at 83 km before and during the NLCs appearance. Arrows show the moments of NLCs observations.*

We see that in all three cases NLCs appeared after the cooling stage, during the hours or day following the temperature minimum provided it is below 150 K (in July, 2015, diurnal variations are also seen). It confirms that this temperature value is a threshold for NLC formation, but a definite amount of time can be required for this process.

**4. NLC scattering field separation**

The main obstacle to NLC ground-based study is bright twilight background consisting of single (basically Rayleigh) scattering in the nearby atmosphere regions and multiple scattering in the troposphere that is complicatedly defined by its current aerosol properties. For scattering angles of 90° and higher (i.e. far from the dusk/dawn segment), the sunlight scattered by NLCs is just a small addition to the surrounding background. It even is during rare events such as July 5, 2015. For spaceborne measurements the problem is less significant but does not completely disappear.

It can seem that we could separate the emission field scattered by NLCs by direct subtraction of twilight background measured on a nearby date at the same depression of the Sun under the horizon in the case of NLCs absence. However, it does not work since the background significantly changes from one twilight to another. Mostly, this is due to variability of the multiple scattering component dominating in the dark twilight stage (Ugolnikov and Maslov, 2007, 2013b).

We could also identify sky points with maximum NLC intensity and subtract the surrounding field. It can work well on some occasions, but here NLCs have no sharp borders (see Figure 2). Note also that NLCs are observed during the transitive and dark twilight stage, according to the classification proposed in (Ugolnikov and Maslov, 2007). In this period the sky background can be also characterized by fast spatial and temporal changes of brightness and polarization due to disappearance/reappearance of a single scattering field. It follows that the background Stokes vector can be significantly different even for the neighboring points of the sky.

Another possible solution is to check the variability of the first and second Stokes vector components of the sky background and to find the synchronous small-scale changes that can be related with NLCs. This way can give better estimates for polarization of light scattered by NLCs than the methods listed above but some problems emerge again. First, three directions of polarization filter during measurements do not coincide with coordinate axes shown in Figure 2b and, therefore, estimates of Stokes parameters for definite sky point and time are not independent. Thus, the polarization value is also contributed by the measurement noise which also gives simultaneous variations of Stokes parameters. Second, if NLCs have no sharp borders, then the spatial changes of brightness and polarization of NLCs and the twilight background can correlate, and that will also affect the result.

Building an optimal scheme for separation of light scattered by NLCs, we take advantage of a large number of measurements (spatially, across the sky, and temporally, during the twilight) and maximally subtract the surrounding background, even if the NLC signal is partially lost as well. We select the single arc with a constant value of scattering angle $\theta$ (Figure 2) and use two basic principles:

1) NLCs are above the layer of multiple scattering and optically thin, and if they appear on the arc, their Stokes vector $\mathbf{B_N}$ should be the simple addition to the sky background $\mathbf{B_S}$ (this is not true, for example, for dense tropospheric clouds). The sunlight scattered by NLCs experiences the extinction in the lower atmospheric layers, but it does not change its polarization properties.

2) We assume that NLCs have uniform characteristics, so that the ratio of Stokes vector components (or the polarization degree) $-B_{N2}/B_{N1}$ is constant along a fixed arc with given angle $\theta$. Real variations are averaged on the observation period during the twilight and contribute to the errors of polarization measurements.

Having measured the total background Stokes vector $\mathbf{B}(\theta, \tau)$ for a definite value of $\theta$ and for $\tau$ in the range from $\tau_{MIN}$ to $\tau_{MAX}$, we can separate the large-scale and short-scale variations of this function by $\tau$:

$$\mathbf{B}(\theta, \tau) = \mathbf{C_0}(\theta) + \mathbf{C_1}(\theta) \cdot \tau + \mathbf{C_2}(\theta) \cdot \tau^2 + \mathbf{b}(\theta, \tau) \qquad (1).$$

Here the vectors $\mathbf{C_0}$, $\mathbf{C_1}$, and $\mathbf{C_2}$ are to be found by the least-squares method. The main contribution to these vectors is due to twilight background, rather than to NLCs, and so they may be discarded from the analysis. The short-scale variations $\mathbf{b}(\theta, \tau)$ are orthogonal to the long-scale ones; that is, they have the property:

$$\int_{\tau_{MIN}}^{\tau_{MAX}} \mathbf{b}(\theta,\tau)\, d\tau = \int_{\tau_{MIN}}^{\tau_{MAX}} \mathbf{b}(\theta,\tau)\tau\, d\tau = \int_{\tau_{MIN}}^{\tau_{MAX}} \mathbf{b}(\theta,\tau)\tau^2\, d\tau = 0 \qquad (2).$$

They are contributed by both NLCs ($\mathbf{b_N}$) and twilight sky background ($\mathbf{b_S}$). Again, we can not subtract $\mathbf{b_S}$ directly, having measured it during the same twilight stage for the nearby date without NLCs (we denote it as $\mathbf{b_0}$). However, we can estimate the fraction of NLCs field variations $\mathbf{b_{NX}}$ orthogonal to those of $\mathbf{b_0}$:

$$\mathbf{b}(\theta, \tau) = \mathbf{D}(\theta)\, \mathbf{b_0}(\theta, \tau) + \mathbf{b_{NX}}(\theta, \tau);$$

$$\int_{\tau_{MIN}}^{\tau_{MAX}} \delta_{ij} b_{NXi}(\theta,\tau) b_{0j}(\theta,\tau)\, d\tau = 0 \qquad (3).$$

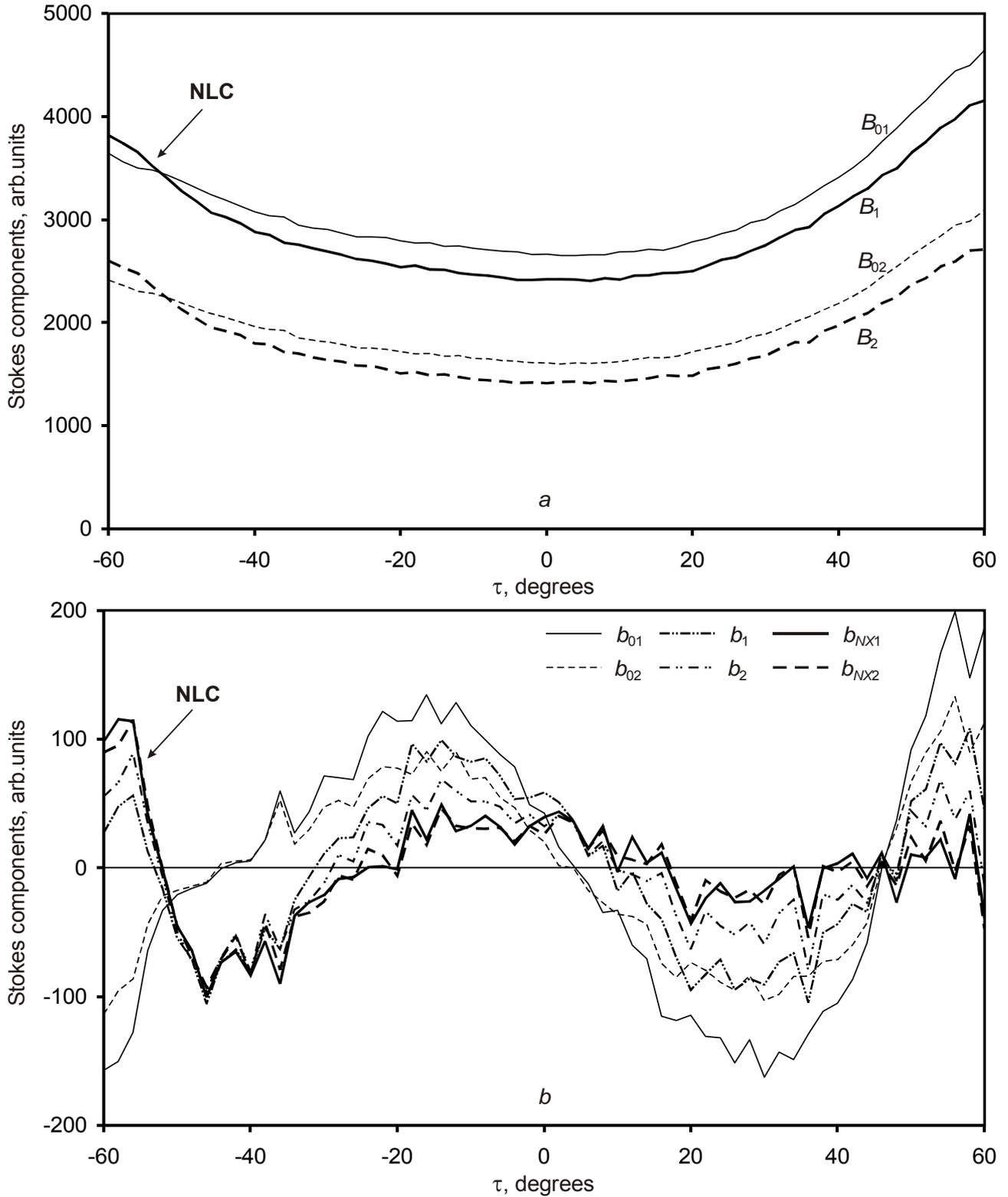

*Figure 4. The example of dependencies of Stokes vector parameters along the arc with $\theta=90°$ (a) and their short-scale variations (b), the morning of July 22, 2014.*

Here **D** is a diagonal matrix to be found by the least-squares method, $\delta_{ij}$ is the Kronecker symbol. This way we completely eliminate all long-scale background variations and also short-scale variations that correlate with those of the reference twilight with no NLCs. The data from measurements made during twilights of morning, July 20, 2014, evening, July 4, 2015 (Chepelevo) and evening, August 19, 2015 (Apatity) at the same solar zenith angles were used as references for three twilights with NLCs listed above, respectively.

Figure 4ab illustrates this procedure to analysis of faint NLCs without distinct edges (July 22, 2014, θ=90°). Dependencies of the first and second components of Stokes vectors $\mathbf{B}(\tau)$ and $\mathbf{B}_0(\tau)$ are plotted in Figure 4a. NLCs appeared in the far left of the dawn ($\tau<-45°$).

Short-scale variations $\mathbf{b}(\tau)$ and $\mathbf{b}_0(\tau)$ are shown in Figure 4b. It is easy to see that they are strongly correlated, and the dominant contribution to the function $\mathbf{b}(\tau)$ comes from the twilight background, which is less polarized than the light scattered by NLCs. Failure to run the orthogonal procedure described by formula (3) results in an underestimate for the polarization value: $-b_2(\tau)/b_1(\tau)\sim0.7$. Orthogonal variations $\mathbf{b}_{NX}(\tau)$ are shown in bold in Figure 4b. They are significantly different from zero only for negative $\tau$, where NLCs appear, and the polarization $-b_{NX2}(\tau)/b_{NX1}(\tau)$ is closer to unity. This procedure works even better in the case of bright NLCs extended along the whole arc (July 5, 2015).

However, the small-scale random noise variations of $b_{NX2}(\tau)$ and $b_{NX1}(\tau)$ remain correlated for the reason described above. To reduce this factor and to maximize accuracy, the value of polarization of the light scattered by NLCs by an angle θ is found as the ratio of variation integrals:

$$p_m(\theta) = -\frac{I_{12}}{I_{11}} = -\frac{\int_{\tau_{MIN}}^{\tau_{MAX}} b_{NX1}(\theta,\tau)\hat{b}_{NX2}(\theta,\tau)d\tau}{\int_{\tau_{MIN}}^{\tau_{MAX}} b_{NX1}(\theta,\tau)\hat{b}_{NX1}(\theta,\tau)d\tau} \quad (4),$$

where

$$\hat{b}_{NXi}(\theta,\tau) = \frac{b_{NXi}(\theta,\tau-\Delta\tau)+b_{NXi}(\theta,\tau+\Delta\tau)}{2} \quad (5).$$

The value of $\Delta\tau$ is equal to the resolution of the data analysis (2°). The resulting value of the polarization $p(\theta)$ for each twilight is obtained by averaging over all measurements:

$$p(\theta) = \frac{\sum_m W_m(\theta)p_m(\theta)}{\sum_m W_m(\theta)} \quad (6).$$

The weights are proportional to the relative contribution of $b_{NX1}$ variations in the total brightness square integral:

$$W_m(\theta) = \frac{\int_{\tau_{MIN}}^{\tau_{MAX}} b_{NX1}(\theta,\tau)\hat{b}_{NX1}(\theta,\tau)d\tau}{\int_{\tau_{MIN}}^{\tau_{MAX}} B_1(\theta,\tau)\hat{B}_1(\theta,\tau)d\tau} \quad (7).$$

The weights help to select the cases with significant appearance of NLCs on an arc. For the same purpose, the following additional data selection criteria were also applied:

1) The solar zenith angle is not less than 96.5°, so NLCs are distinct on the twilight background;
2) The effective baseline altitude of single scattering (or effective shadow altitude, Ugolnikov and Maslov, 2013b) in the middle of the arc is not more than 75 km, so NLCs are well sunlit;
3) The correlation of $b_{NX1}(\tau)$ and $\hat{b}_{NX2}(\tau)$ values is more than 0.9 (after that we eliminated the noise auto-correlation).

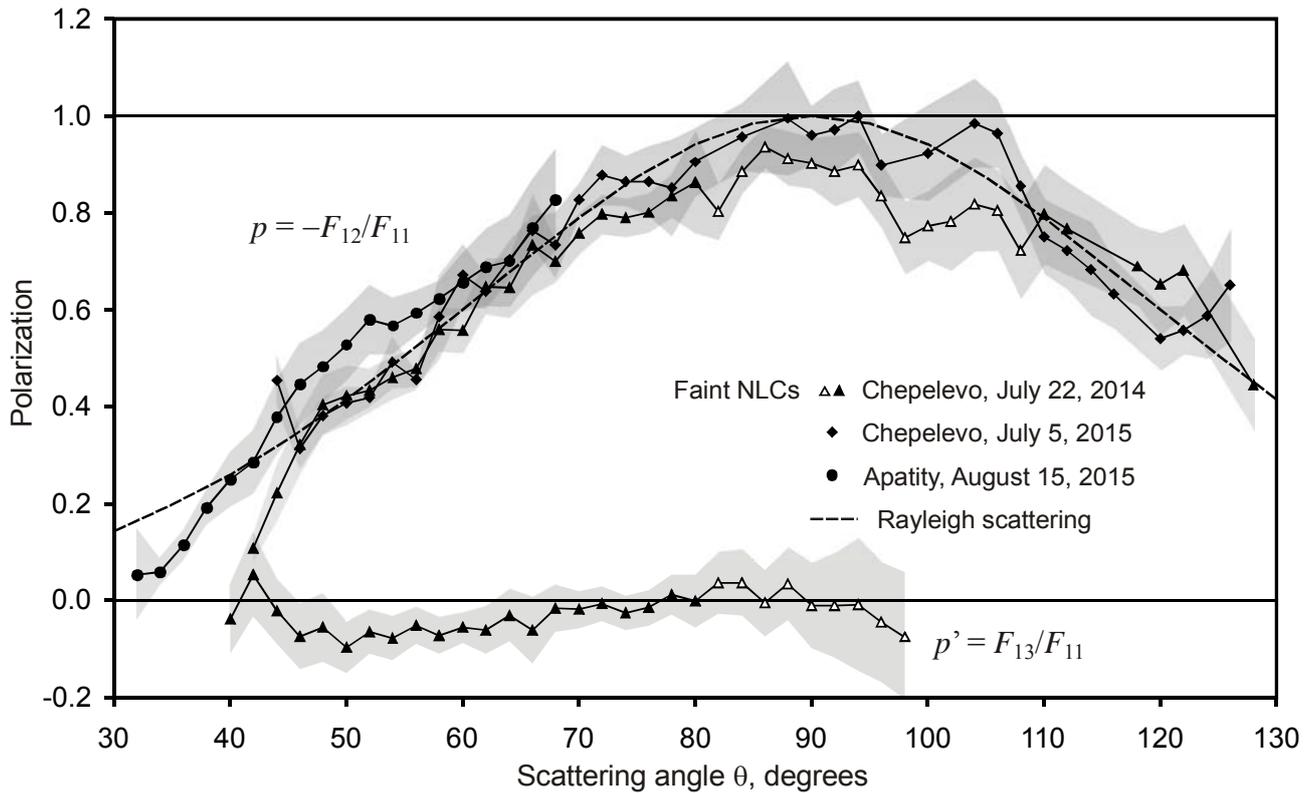

*Figure 5. The observed scattering angle dependences of the polarization of the light scattered by NLCs. Errors are shown in grey, p and p' values are described in the text.*

This procedure was applied separately for different θ values with resolution 2° to build the final $p(\theta)$ dependencies.

**5. NLCs polarization analysis**

The results of the procedure described above are shown in Figure 5, where points with averaging errors (6) of less than 0.12 are plotted. The Rayleigh scattering dependence for tiny spheres is also drawn. Principal agreement of the Rayleigh model with observations is clearly seen (taking measurement errors into account). There are two possible reasons for systematic differences between measured and real dependencies $p(\theta)$:

1) Incomplete subtraction of the twilight background;
2) Scattering of the atmospheric background by NLCs, i.e. multiple scattering in the atmosphere with the last scattering on the NLC particle.

Note that the first problem arises only for ground-based analysis, and the second one is common for ground-based, rocketborne and satellite measurements. Influence of the non-subtracted background is expected to increase in the case of faint NLCs, so it should lead to different polarization values for bright and faint NLCs. Comparing the data of July 22, 2014 and of July 5, 2015 for the same location, we see a polarization decrease in the morning of July 22, 2014 at the scattering angles 80-110°, where NLCs were especially faint (see Figure 2a, and the open triangles in Figure 5). For all other scattering angles there is a good agreement between two twilights data showing no significant effects due to non-subtracted background.

The problem of multiple scattering contribution is more complicated and less dependent on the brightness of NLCs. Obviously, multiple scattering should decrease the measured degree of polarization of the light scattered by NLCs. The values close to unity obtained on July 5, 2015 for the scattering angles around 90° indicate that such an effect is negligibly small, at least for these angles. To check other angles, we take into account that effective position of secondary light source for NLCs (atmospheric background) does not coincide with the Sun and that measurement of the third component $F_{13}$ of the NLCs Stokes vector does not generally give zero result. The same effect can be used for the separation of single and multiple scattering in the twilight analysis (Ugolnikov and Maslov, 2015).

The effect of non-zero third component vanishes in the solar vertical ($\tau=0$) owing to symmetry of the radiation transfer scheme. The sign of the third component $p'$ is different eastwards and westwards from this line in the sky. The best measurement we have for its estimation is from observation of the morning twilight of July 22, 2014, when NLCs appeared only on one side of the solar vertical, far from it ($\tau<-45°$). We can run the procedure described in Section 4 replacing the second Stokes component by the third one but not requiring good correlation between $b_{NX1}(\tau)$ and $\hat{b}_{NX3}(\tau)$ values.

The results have good accuracy for scattering angles in the range 40° to 100° (see Figure 5). The value of $p'$ is insignificant but exceeds the measurement error for $\theta<60°$. It has the same sign as the twilight background close to NLCs ($\tau<0$). Another possible property of multiple scattering is a fast decrease of the polarization of the light scattered by NLCs, $p$, compared to the Rayleigh value at $\theta<50°$ (see Figure 5). Earlier observations (Vasilyev, 1959, 1967) showed that $p$ could even take negative values for lower scattering angles. The same situation is observed for the multiple scattering component of the twilight background: the "inverted polarization" effect appears closer to the Sun (Ugolnikov, 1999). The non-subtracted background can also contribute to this drop in degree of polarization for small scattering angles as it was observed on July 22, 2014 but not on July 5, 2015.

The above consideration shows that multiple scattering (or background scattering on NLC particles) does not crucially affect the measured polarization value, especially for scattering angles close to 90° which are most important for analysis. This is confirmed by high values of polarization ($p\sim1$) on the evening of July 5, 2015. However, multiple scattering cannot be completely neglected in the dusk/dawn area (for small scattering angles). Far from the solar vertical this effect can give rise to small non-zero third and, consequently, fourth component of the Stokes vector for NLCs (Bohren, 1983). Note that the circular polarization of NLCs was detected by Gadsden et al. (1979) at about 25° from the solar vertical. We believe that a strong indication of the circular polarization of light scattered by NLCs would be its detection right above the Sun in the sky or its observation in the UV-range of spectrum, where the multiple scattering contribution is lower due to absorption by stratospheric ozone.

## 6. Particle size and shape analysis

Dependencies of degree of polarization of the light scattered by NLCs shown in Figure 5 are in good agreement with the Rayleigh values for different shapes (van de Hulst, 1957), so the assumption of very small particles is principally confirmed by the observation results. Comparing them with the theoretical data for larger particles, we can estimate their most probable (or best-fit) and maximum sizes. Note that in the case of the sharp-edged size particle distribution, largest particles give a dominant contribution to the optical NLC properties (Gadsden, 1981) and monodisperse models can be effective for such a comparison. However, the maximum size value can be considerably different for spherical and non-spherical particles (Gadsden, 1978; Bohren, 1983; Mishchenko, 1992) because of the different polarization properties of the scattered light.

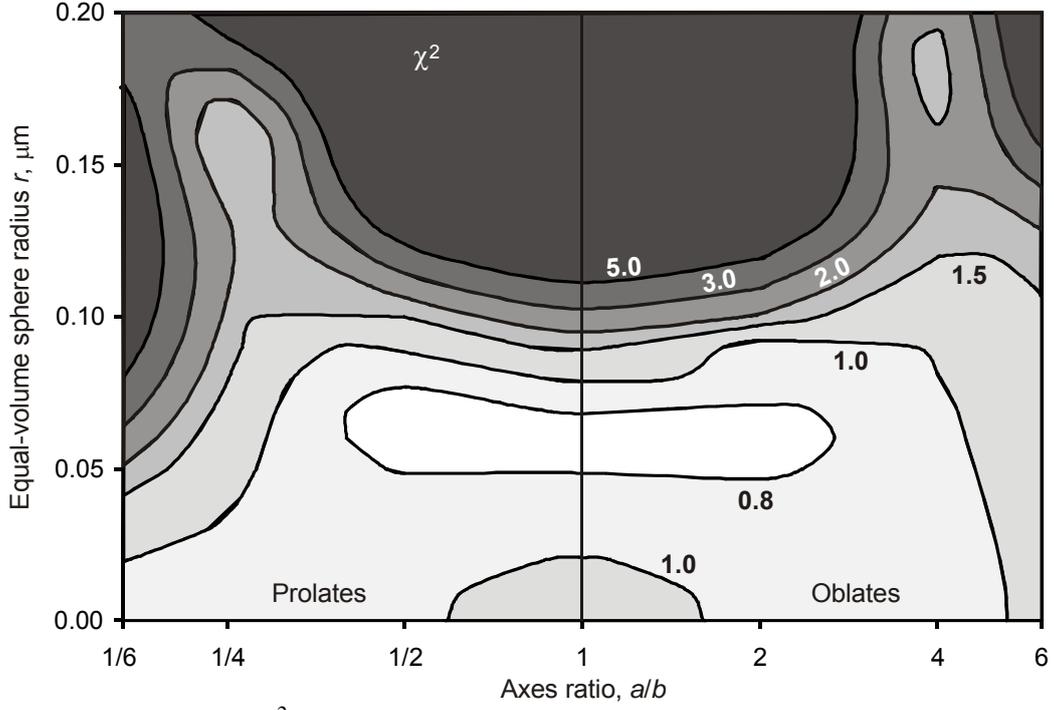

*Figure 6. The distribution of $\chi^2$ value on the size and shape of ice particles.*

Checking the agreement of the experimental data with different particle models, we take the value of ice refractive index equal to 1.31. Following Iwabuchi and Yang (2011), we do not assume it to depend on temperature in optical spectral range. Ice optical properties are not well-known for low temperatures (Kokhanovsky, 2005). The laboratory measurements of cold ice films (Westley et al., 1998) give the value 1.28±0.02 for 140 K, but actually the refractive index is found out to be independent on temperature again. As authors note, there is the systematic offset of about 0.03 related with less density of ice films compared with ice crystals. If we take it into account, then we return to the value 1.31 for NLCs.

We calculate the dependencies of Stokes vector component ratio $p_T = -F_{12}/F_{11}(\theta)$ for ice spheroids with equal-volume sphere radii in the range of possible values for NLCs up to 0.20 μm. Models of prolate and oblate spheroids with axes ratio $a/b$ from 1/6 to 6 (Mishchenko, 1992) were used. We also use Waterman's T-matrix approach (Mishchenko et al., 1996) and the FORTRAN code developed by M.I. Mishchenko (publicly available at http://www.giss.nasa.gov/staff/mmishchenko/t_matrix.html).

The basic criterion for model consistency involves the value of

$$\chi^2 = \frac{1}{\sqrt{2N}} \left( \sum_{n=1}^{N} \frac{(p(\theta_n) - p_T(\theta_n))^2}{\sigma^2(\theta_n)} - N \right) \quad (8),$$

where $p(\theta_n)$ and $p_T(\theta_n)$ stand for observational and theoretical degrees of polarization at the scattering angle $\theta_n$, $\sigma$ is the error of measurements, $N$ is the total number of scattering angle values. To avoid the systematical errors discussed above, we use the $p$ and $\sigma$ values of July 5, 2015 with bright all-sky NLCs and minimal contribution of the sky background as the observational data. This is the only reliable twilight polarization data in the most important interval at $\theta \geq 90°$ and without the essential effects of a decrease in $p$ for small scattering angles due to the multiple scattering or non-subtracted background (see Figure 5).

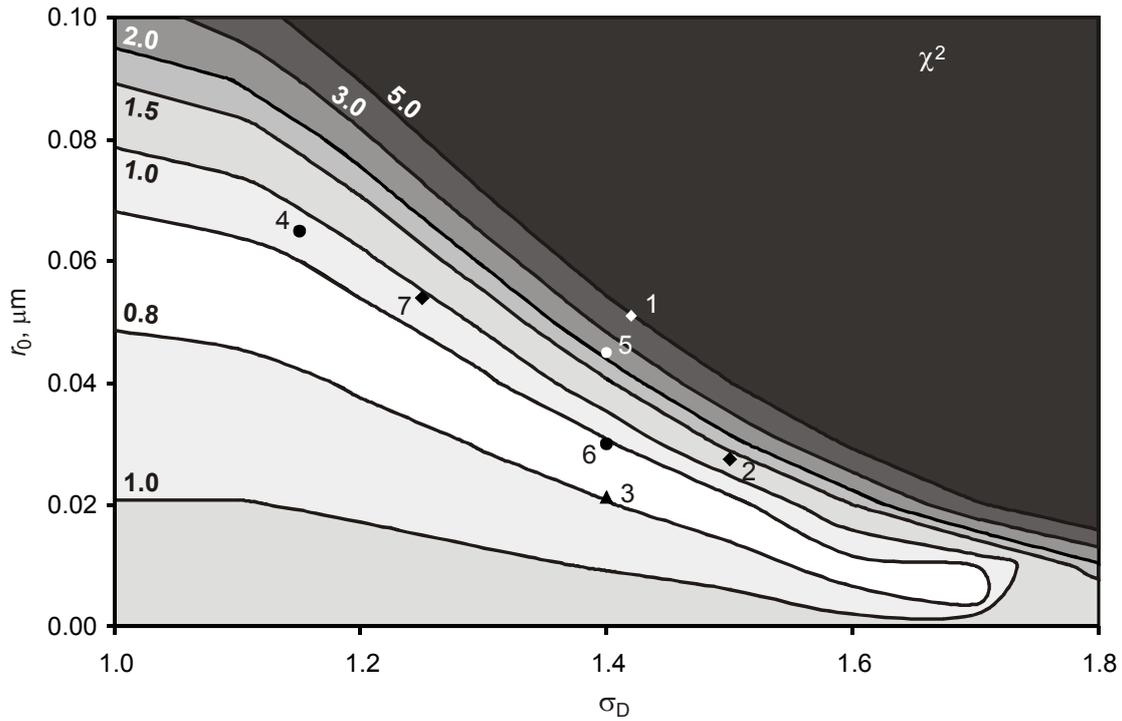

*Figure 7. The $\chi^2$ value depending on the parameters of log-normal distribution of spherical ice particles on size. The points correspond to the previous lidar (diamonds), rocketborne (triangle) and spaceborne (circles) measurements listed by Kokhanovsky (2005) and two more later ones, see references below Eq.9.*

If the model is in a good agreement with observations then $\chi^2$ will be less than or close to unity, and this is true for the Rayleigh model of scattering, where $\chi^2=1.06$. Models with $\chi^2 \geq 2$ are considered to be less probable while those with $\chi^2 \geq 3$ are rejected outright. Figure 6 shows the dependence of $\chi^2$ on the particles size and shape for monodisperse spheres and spheroids. The most probable NLC particle radius corresponding to the minimum of $\chi^2$ is 0.06 μm being in agreement with other methods of measurements (Gumbel et al., 2001; Kohanovsky, 2005, and references therein). Spherical or weakly elongated particles (with axes ratio from 1/2 to 2) provide the best-fit between theory and observations.

To describe the distribution of NLC particles with respect to size, the log-normal polydisperse model is widely used, see e.g. Kokhanovsky (2005):

$$f(r) = \frac{1}{\sqrt{2\pi}\zeta r} \exp\left(\frac{-\ln^2(r/r_0)}{2\zeta^2}\right); \quad \zeta = \ln \sigma_D \tag{9}$$

Here $r_0$ and $\sigma_D$ are free parameters of the model. Distribution (9) goes into the monodisperse in the limit $\sigma_D=1$. Integrating the Stokes vector phase functions calculated for particles with different sizes, we can also check these models by using formula (8). In the case of spherical particles the results are shown in Figure 7. We compare them with the previous results of lidar, rocketborne and spaceborne measurements represented respectively by diamonds, triangle and circles (1 – Von Cossart et al., 1999; 2 – Alpers et al., 2000; 3 – Gumbel and Witt, 2001; 4 – Carbary et al., 2002; 5 – Von Savigny et al., 2004; 6 – Von Savigny and Burrows, 2007, for close latitudes; 7 – Baumgarten et al, 2007).

Evidently, $r_0$ and $\sigma_D$ are related in the sense that only the less mean radius $r_0$ is possible for widely polydisperse models (the more $\sigma_D$). Otherwise, the particles with size above 0.1 μm will appear, that will increase the $\chi^2$ value. Given $\sigma_D=1.4$ (the standard value used in many papers), the most probable value of $r_0$ by polarization measurements will be about 0.025 μm, which is in best agreement with the data of rocketborne (Gumbel and Witt, 2001) and spaceborne (Von Savigny and Burrows, 2007) measurements.

The maximal size is strongly dependent on particle shape. Assuming a spherical shape for all particles and using Mie theory we obtain a maximum radius of about 0.1 μm which is less than that for spheroids. Models of prolate (or oblate) spheroids with axes ratio equal to 1/4 (or 4) result in the function $p(\theta)$ being symmetric with maximum near to $\theta=90°$ and turn out to be in satisfactory agreement with observations for significantly higher sizes of up to 0.2 μm (see Figure 6). This is also seen in Figure 8, where the functions $p(\theta)$ near to local minima of $\chi^2$, including the above two types of spheroids, are compared to the observational data from July 5, 2015. We also note that asymmetrical particles with close axes ratios were proposed by Baumgarten et al. (2002) as an explanation of the depolarization effect in NLCs lidar sounding.

## 7. Conclusion

We analyzed measurements of degree of polarization due to sunlight scattering by noctilucent clouds in a wide range of scattering angles, including the back hemisphere ($\theta>90°$). This is important for two reasons: first, the polarization becomes more sensitive to the size and shape of particles at these angles, and second, the effect of size increase (at least for spherical and weakly elongated particles) causes a drift of the maximum polarization towards large angles. As a result the influence of such an effect is different from those of the multiple scattering or non-subtracted sky background. For small scattering angles all effects lead to a homogenous decrease in the measured polarization and can be easily mixed up with each other.

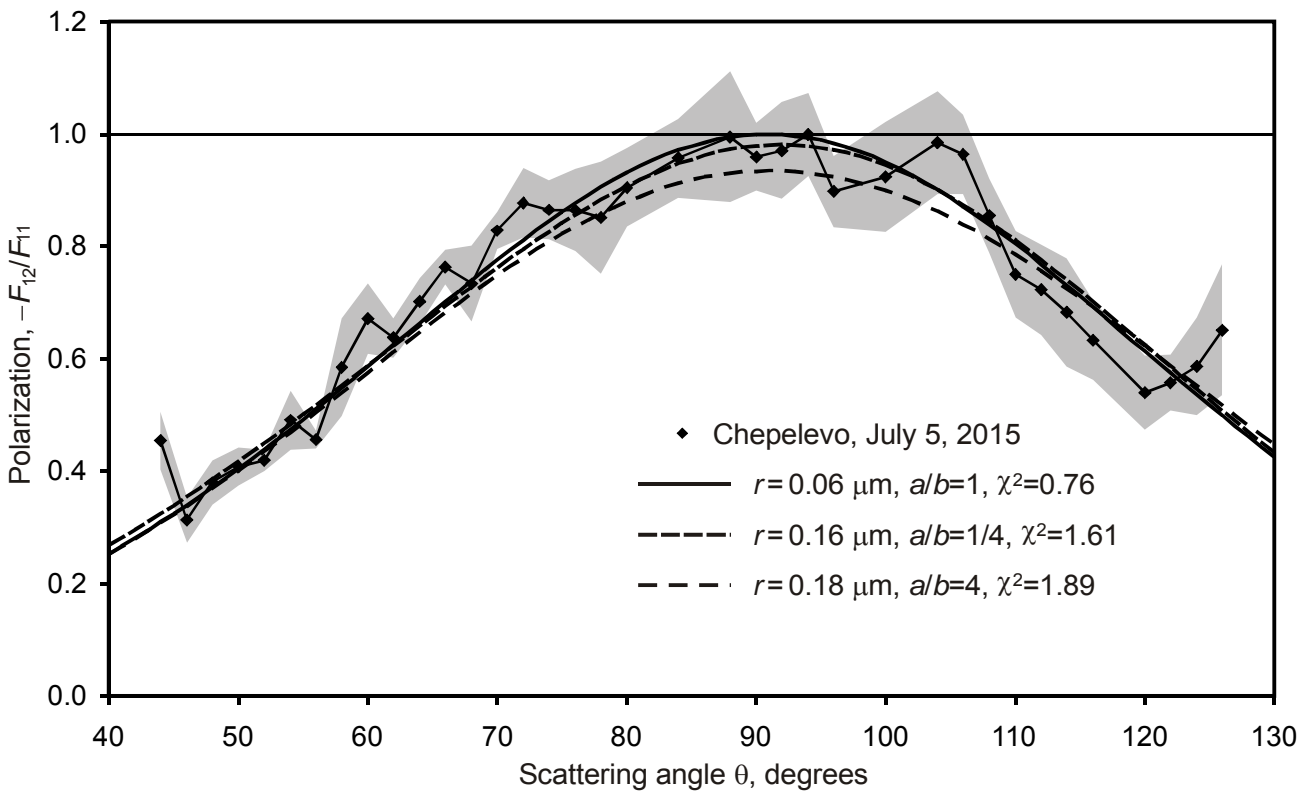

*Figure 8. Polarization of the light scattered by NLCs in the evening of July 5, 2015, compared with monodisperse models of ice spheres and spheroids near the local minima of $\chi^2$ distribution in Figure 6.*

However, NLCs are quite faint and are rarely visible in the sky hemisphere corresponding to large scattering angles; that is opposite to the Sun. The accuracy of single point measurement can not be high, but integrating along an arc with constant θ and averaging over the large number of fast CCD measurements made during twilight (especially, at high latitudes) can solve this problem. Focusing on short-scale background variations orthogonal to the ones of the twilight sky without NLCs, we significantly decrease the influence of twilight background to the NLC polarization analysis.

The Rayleigh-type dependence of the polarization of light scattered by NLCs indicates small sizes of the particles. The best-fit monodisperse value of the radius is 0.06 μm, and the $r_0$ value for the log-normal distribution with $\sigma_D$=1.4 is 0.025 μm. It is worth noting that these estimates are in good agreement with other measurements and were made by using of one-color analysis with a wavelength 10 times larger than particle radius. Obviously, the accuracy can be improved provided the observations are done in several wavelengths including shorter ones.

However, the maximal size value sufficiently depends on the particles shape. If we are sure that strongly elongated ice crystals can not appear in NLCs (Turco et al., 1982; Kokhanovsky, 2005), then we will have $r$<0.1 μm, which is a stronger restriction than was evident according to previous measurements of the polarization of the light scattered by NLCs in the visible part of spectrum (only the range θ<90° was analyzed in those papers). However, as shown by Baumgarten et al. (2002) and Rapp et al. (2006), elongated particles can exist and, therefore, their sizes can be larger than those for spherical particles.

A more exact estimate of NLC particle sizes can be achieved by the use of polarimetry in the blue and ultraviolet part of spectrum, the latter is possible in rocket or space measurements. However, to increase the accuracy and to fix the possible contribution of multiple scattering, NLCs polarimetry must also include large scattering angles. This contribution (as for the twilight sky) reaches its maximum in the blue and violet spectral regions (Ugolnikov and Maslov, 2002), then decreases (but does not vanish) in ultraviolet range due to ozone absorption. The multiple scattering effects can be the cause of circular polarization of the light scattered by NLCs but in this case it should vanish in the solar vertical, and this needs to be checked in future observations.

**Acknowledgments**
We are grateful to Yury S. Ivanov (Main Astronomical Observatory of the National Academy of Sciences of Ukraine) for general optical scheme idea and providing the main lens used in all-sky polarization measurements in Apatity, Michael I. Mishchenko (NASA Goddard Institute of Space Science, USA) for providing the T-matrix method code, Gerd Baumgarten (Leibniz-Institute of Atmospheric Physics, Kühlungsborn, Germany) for useful remarks. We also thank Konstantin B. Alkalaev and Ian Marshall for their help in manuscript preparation.

The work is supported by Russian Foundation for Basic Research, grant No. 16-05-00170.

**References**

Alpers, M., Gerding, M., Höffner, J., von Zahn, U., 2000. NLC particle properties from a five-color lidar observation at 54°N. J. Geophys. Res. 105 (D10), 12235-12240.

Baumgarten, G., Fricke, K.H., von Cossart, G., 2002. Investigation of the shape of noctilucent cloud particles by polarization lidar technique. Geophys. Res. Lett. 29, 1-4.

Baumgarten, G., Fiedler, J., von Cossart, G., 2007. The size of noctilucent cloud particles above ALOMAR (69N,16E): Optical modeling and method description. Adv. Space Res. 40, 772-784.

Berger, U., Lübken, F.J., 2015. Trends in mesospheric ice layers in the Northern Hemisphere during 1961–2013. J. Geophys. Res. 120, 11277-11298.

Bohren, C.F., 1983. On the size, shape and orientation of noctilucent cloud particles. Tellus 35B, 65-72.

Bronshten, V.A., Grishin, N.I., 1970. Noctilucent clouds. Nauka, Moscow (in Russian).

Carbary, J.F., Morrison, D., Romick, G.J., 2002. Particle characteristics from the spectra of polar mesospheric clouds. J. Geophys. Res. 107 (D23), 4686.

Dalin, P., Kirkwood, S., Andersen, H., Hansen, O., Pertsev, N., Romejko, V., 2006. Comparison of long-term Moscow and Danish NLC observations: statistical results. Ann. Geophys., 24, 2841–2849.

Dalin, P., Pertsev, N., Dubietis, A., Zalcik, M., Zadorozhnyi, A., Connors, M., Schofield, I., McEvan, T., McEachran, I., Frandsen, S., Hansen, O., Andersen, H., Sukhodoev, V., Perminov, V., Balciunas, R., Romejko V., 2011. A comparison between ground-based observations of noctilucent clouds and Aura satellite data. J. Atm. Sol-Terr. Phys 73, 2097-2109.

DeLand, M.T., Thomas, G.E., 2015. Updated PMC trends derived from SBUV data. J. Geophys. Res. 120, 2140-2166.

Donahue, T.M., Guenther B., Blamont, J.E., 1972. Noctilucent clouds in daytime circumpolar particulate layers near the summer mesopause. J. Atmos. Sci. 29, 1205-1209.

Donahue, T.M., Guenther, B., 1973. The altitude of the scattering layer near the mesopause over the summer pole. J. Atmos. Sci. 30, 515-517.

EOS MLS Science Team, 2011. MLS/Aura Level 2 Temperature,version 003, Greenbelt, MD, USA:NASA Goddard Earth Science Data and Information Services Center, Accessed Enter User Data Access Date at http://disc.sci.gsfc.nasa.gov/datacollection/ML2T_V003.html

Gadsden, M., 1978. The size of particles in noctilucent clouds: implications for mesospheric water vapor. J. Geophys. Res. 83, 1155-1156.

Gadsden, M., Rothwell, P., Taylor, M.J., 1979. Detection of circularly polarized light from noctilucent clouds. Nature 278, 628-629.

Gadsden, M., Schröder, W., 1989. Nocilucent Clouds. Springer-Verlag, Berlin.

Gumbel, J., Stegman, J., Murtagh, D.P., Witt, G., 2001. Scattering phase functions and particle sizes in noctilucent clouds. Geophys. Res. Lett. 28, 1415-1418.

Gumbel, J., Witt, G., 2001. Rocket-borne photometry of NLC particle populations. Adv. Space Res. 28, 1053-1058.

Heitzenberg, J., Witt, G., Kinmark, I., 1978. Optical characteristics of noctilucent clouds: measurements and interpretation. Proc. 6$^{th}$ Ann. Meet. Upper Atmos. Stud. Opt. Methods, 78-84.

Hemenway, C.L., Hallgren, D.S., 1969. Collection of meteoric dust after the Leonid meteor shower 1965. Space Research IX, 140-146.

Hervig, M., Thompson, R.E., McHugh, M., Gordley, L.L., Russell, G.M. III, 2001. First confirmation that water ice is the primary component of polar mesospheric clouds. Geophys. Res. Lett. 28, 971-974.

Hunten, D.M., Turco, R.P., Toon O.B., 1980. Smoke and Dust Particles of Meteoric Origin in the Mesosphere and Stratosphere. J. Atmos. Sci. 37, 1342-1357.

Iwabuchi, H., Yang, P., 2011. Temperature dependence of ice optical constants: Implications for simulating the single scattering properties of cold ice clouds. J. Quant. Spectrosc. Radiat. Transfer, 112, 2520-2525.

Kokhanovsky, A.A., 2005. Microphysical and optical properties of noctilucent clouds. Earth-Science Reviews 71, 127-146.

Leslie, R.C., 1885. Sky glows. Nature 32, 245.

Mishchenko, M.I., 1992. Light scattering by nonspherical ice grains: an application to noctilucent cloud particles. Earth, Moon, Planets 57, 203-211.

Mishchenko, M.I., Travis, L.D., Mackowski, D.W., 1996. T-matrix computations of light scattering by nonspherical particles: A review. J. Quant. Spectrosc. Radiat. Transfer 55, 535-575.

Pertsev, N., Dalin, P., Perminov, V., Romejko, V., Dubietis, A., Balčiunas, R., Černis, K., Zalcik, M., 2014. Noctilucent clouds observed from the ground: sensitivity to mesospheric parameters and long-term time series. Earth, Planets and Space, 66, 98.

Rapp, M., Thomas, G.E., 2006. Modeling the microphysics of mesospheric ice particles: assessment of current capabilities and basic sensitivities. J. Atmos. Solar Terr. Phys. 68, 715–744.

Rapp, M., Thomas, G.E., Baumgarten, G., 2006. Spectral properties of mesospheric ice clouds: Evidence for nonspherical particles. J. Geophys. Res. 112, D03211.

Roble, R.G., Dickinson, R.E., 1989. How will changes in carbon dioxide and methane modify the mean structure of the mesosphere and thermosphere? Geophys. Res. Lett. 16, 1441-1444.

Romejko, V.A., Dalin, P.A., Pertsev, N.N., 2003. Forty years of noctilucent cloud observations near Moscow: database and simple statistics, J. Geophys. Res., 108(D8), 8443.

Rosinski, J., Snow, R.H., 1961. Secondary particulate matter from meteor vapors. J. Meteorol. 18, 736–745.

Russell, J.M. III, et al., 2009. The Aeronomy of Ice in the Mesosphere (AIM) mission: Overview and early science results. J. Atmos. Solar Terr. Phys. 71, 289-299.

Schwartz, M.J., et al., 2008. Validation of the Aura Microwave Limb Sounder temperature and geopotential height measurements. J. Geophys. Res. 113, D15S11.

Skrivanek, R.A., Chrest, S.A., Carnevale, R.F., 1969. Particle collection results from recent rocket and satellite experiments. Space Research IX, 129-139.

Soberman, R.K., 1969. The extraterrestrial origin of noctilucent clouds. Space Research IX, 183-189.

Tarasova, T.M., 1962. The polarization of light from noctilucent clouds. Proceedings of the conference on noctilucent clouds, III, Tallinn, 16-19 May 1961, the Academy of Sciences of the Estonian S.S.R., 55-67.

Thomas, G.E., Olivero J., 2001. Noctilucent clouds as the possible indicators of global change in the mesosphere. Adv. Space Res. 28, 937-946.

Tozer, W.F., Beeson, D.E., 1974. Optical model of noctilucent clouds based on polarimetric measurements from two sounding rocket campaigns. J. Geophys. Res. 79 (36), 5607-5612.

Turco, R.P., Toon, O.B., Whitten, R.C., Keesee, R.G., Hollenbach, D., 1982. Noctilucent clouds: simulation studies of their genesis, properties and global influences. Plan. Space Sci. 30, 1147-1181.

Ugolnikov, O.S., 1999. Twilight Sky Photometry and Polarimetry: The Problem of Multiple Scattering at the Twilight Time. Cosmic Research. 37, 159-166.

Ugolnikov, O.S., Maslov, I.A., 2002. Multicolor Polarimetry of the Twilight Sky. The Role of Multiple Light Scattering as a Function of Wavelength. Cosmic Research. 40, 224-232.

Ugolnikov, O.S., Maslov I.A., 2007. Detection of Leonids meteoric dust in the upper atmosphere by polarization measurements of the twilight sky. Plan. Space Sci. 55, 1456-1463.

Ugolnikov, O.S., Maslov, I.A., 2013a. Undisturbed Mesosphere Optical Properties from Wide-Angle Frequent Twilight Sky Polarimetry. Cosmic Research 51, 235-240.

Ugolnikov, O.S., Maslov, I.A., 2013b. Summer mesosphere temperature distribution from wide-angle polarization measurements of the twilight sky, J. Atmos. Solar Terr. Phys. 105-106, 8-14.

Ugolnikov, O.S., Maslov, I.A., 2014. Mesosphere light scattering depolarization during the Perseids activity epoch by wide-angle polarization camera measurements. Plan. Space Sci. 92, 117-120.

Ugolnikov, O.S., Maslov, I.A., 2015. Analysis of twilight background polarization directions across the sky as a tool for multiple scattering separation. Cosmic Research (in press); e-print arxiv.org/pdf/1503.01635.pdf.

Ugolnikov, O.S., Kozelov, B.V., 2015. Mesosphere study by wide-field twilight polarization measurements: first results beyond the Polar Circle. Cosmic Research (in press); e-print arxiv.org/pdf/1504.07353.pdf.

Van de Hulst, 1957. Light scattering by small particles. Wiley, New York.

Vasilyev, O.B., 1959. The experiment on photometric observations of noctilucent clouds, Proceedings of the conference on noctilucent clouds, Tartu, 12-14 December 1958, the Academy of Sciences of the Estonian S.S.R., 77-84.

Vasilyev, O.B., 1962. Results of the absolute photometry and polarimetry of noctilucent clouds, Proceedings of the conference on noctilucent clouds, III, Tallinn, 16-19 May 1961, the Academy of Sciences of the Estonian S.S.R., 29-54.

Vasilyev, O.B., 1967. Astrophysical investigations of noctilucent clouds. Astrosovet, Academy of Sciences of USSR (in Russian).

Von Cossart, G., Fiedler, J., von Zahn, U., 1999. Size distributions of NLC particles as determined from 3-color observations of NLC by ground-based lidar. Geophys. Res. Lett. 26, 1513-1516.

Von Savigny, C., Kokhanovsky, A.A., Bovensmann, H., Eichmann, K.-U., Kaiser, J., Noël, S., Rosanov, A.V., Skupin, J., Burrows, J.P., 2004. NLC detection and particle size determination: first results from SCIAMACHY on ENVISAT. Adv. Space Res. 34, 851-856.

Von Savigny, C., Burrows, J.P., 2007. Latitudinal variation of NLC particle radii derived from northern hemisphere SCIAMACHY/Envisat limb measurements. Adv. Space Res. 40, 765-771.

Westley, M.S., Barrata, G.A., Baragiola, R.A., 1998. Density and index of refraction of water ice films vapor deposited at low temperatures. J. Chem. Phys. 108, 3321-3326.

Whipple, F.L., 1950. The theory of micro-meteorites. Part I. In an isothermal atmosphere. Proc. Nat. Acad. Sci. 36, 687-695.

Willmann, Ch. J., 1962. On the polarization of light from noctilucent clouds, Proceedings of the conference on noctilucent clouds, III, Tallinn, 16-19 May 1961, the Academy of Sciences of the Estonian S.S.R., 29-54.

Witt, G., 1957. Noctilucent clouds observations. Tellus, 9, 3.

Witt, G., 1960. Polarization of light from noctilucent clouds. J. Geophys. Res. 65 (3), 925-933.

Witt, G., 1969. The nature of noctilucent clouds. Space Research IX, 157-169.

Witt, G., Dye, J.E., Wilhelm, N., 1976. Rocket-borne measurements of scattered sunlight in the mesosphere. J. Atmos. Terr. Phys. 38, 223-238.